# Inter-seasonal and multi-objective optimization of a sustainable hydrogen supply chain in Corsica integrating water availability constraints

[a]T. Moustapha Mai[*], [b]C. Azzaro-Pantel, [a]M. Chin Choi, [a]M. Hajajji, [a]C. Cristofari

[a]UMR CNRS 6134 Scientific Centre Georges Peri, University of Corsica, Route des Sanguinaires, F20000 Ajaccio, France,

e-mail:moustapha-Mai_m@univ-corse.fr[*]; e-mail:cristofari@univ-corse.fr; e-mail:chin-choi_m@univ-corse.fr; e-mail: hajjaji_m@univ-corse.fr

[b]Laboratoire de Génie Chimique, Université de Toulouse, CNRS, INPT, UPS, France,

e-mail:catherine.azzaropantel@toulouse-inp.fr

## Abstract

This study investigates the potential of hydrogen as a sustainable energy carrier for mobility applications in island territories, which are traditionally dependent on fossil fuel imports. Green hydrogen is identified as a key component of the energy transition. A Mixed Integer Linear Programming (MILP) model with a multi-period, multi-objective framework is used to optimize the hydrogen supply chain based on system costs, greenhouse gas (GHG) emissions, and a risk index. The model incorporates critical island-specific factors such as water resource availability, renewable energy sources, tourism flow, and geographic constraints. A multi-criteria decision-making tool based on a modified version of TOPSIS (Technique for Order Preference by Similarity to Ideal Solution) aids the identification of optimal solutions. Results suggest a decentralized Hydrogen Supply Chains (HSC) structure with minimized transport. The levelized cost of hydrogen (LCOH) is estimated at 6.54 €/kg, and GHG emissions range from 1.32 to 1.75 kgCO$_2$e/kg H$_2$. This study highlights the impact of tourism on energy demand and the crucial role of water resources, offering a novel approach to optimizing island-specific HSC.

**Highlights**

Hydrogen as a sustainable energy carrier for island mobility, replacing fossil fuel imports.
MILP model optimizes HSC considering costs, GHG emissions, and risk index.
Island-specific factors like water availability, renewable energy, tourism, and geography are incorporated.
Decentralized HSC minimizes transport costs, with LCOH at 6.54 €/kg.
Tourism impact and water resource constraints play a crucial role in optimizing hydrogen systems.

---

[*] Tchougoune Moustapha Mai
moustapha-Mai_m@univ-corse.fr

# Abbreviation

| | |
|---|---|
| *ADEME* | French environment and energy management agency |
| *AEL* | Alkaline Electrolyzer |
| *BEVs* | Battery electric vehicles |
| *CAPEX* | Capital Expenditure |
| *EDF* | French energy supplier (Electricité de France) |
| *El* | Electrolyzer |
| *f* | Scale factor |
| *FC* | Fuel Cell |
| *FCEV* | Fuel Cell Electrical Vehicle |
| *GHG* | Greenhouse gas |
| *GIS* | Geographic Information System |
| *HSC* | Hydrogen Supply Chain |
| *LCA* | Life Cycle Analysis |
| *LCOH* | Levelized cost of hydrogen |
| *MILP* | Mixed Integer Linear Programming |
| *m-TOPSIS* | Modified Technique for Order Preference by Similar to Ideal Solution |
| *NIAs* | Non-Interconnected Areas |
| *OEHC* | Corsican Hydraulic Equipment Office |
| *OPEX (O&M)* | Operating Expenditure |
| *PEM* | Proton Exchange Membrane |
| *PPE* | French Energy Plan |
| *PV* | Photovoltaic energy |
| *SO* | Solid Oxide |
| *TDC* | Total Daily Cost |
| *Wind* | Wind Energy |

## *Sets*

| | | |
|---|---|---|
| *e* | Energy sources type | *Photovoltaic, Wind, Grid* |
| *fs* | Refuelling stations technology | *Gas supply station, Liquide supply station* |
| *g (or g')* | Grids distribution | *[1,…,9]* |
| *i* | Hydrogen physical form | *Gas, liquid* |
| *j* | Facility size | *Mini, small, medium, large* |
| *l* | Transportations modes | *Tube trailer, tanker truck* |
| *m* | Time discretization, i.e. monthly period | *[1,…,12]* |
| *p* | Production plant technology (electrolysis) | *PEM, AE* |
| *s* | Storage plant technology | *Gas storage, liquid storage* |
| *t* | Time discretization, i.e. yearly period | *[1,…,6]* |

## *Parameters*

| | |
|---|---|
| $AD_{g,g'}$ | Average distance between grids *(km)* |
| $CleanWater_t$ | Total amount of potable water distributed in Corsica (Mm$^3$) |
| *dr* | Discounted rate (%) |
| $DT_{gtm}$ | Total hydrogen demand *(kg/day)* |
| $eCo_{te}$ | Capacity evolution rate of energy sources |
| $ELCF^{max}_{p,j,i}$ | Maximum capacity factor of electrolyser |
| $ELCF^{min}_{p,j,i}$ | Minimum capacity factor of electrolyser |
| *ELWUC* | Water unit consumption by electrolyser (l/kg) |
| *er* | Margin of error *(%)* |
| *Etoe* | Conversion factor of one tone of oil equivalent (toe) into kilowatt-hours (kWh) |
| *f* | Inflation rate *(%)* |
| *FCEVCons* | Hydrogen consumption rate by FCEV *(kg/100km)* |
| *FCmar* | Fuel consumption related to goods transportation (toe) |
| *FCres* | Fuel consumption related to resident transportation (toe) |
| *FCtour* | Fuel consumption related to tourist transportation (toe) |
| *FHV* | Lower heating value of hydrogen (kWh/kg) |
| $\gamma_{pj}$ | Rate of utilisation of primary energy source *(kwh/kg)* |
| *Gtour$_g$* | Concentration rate of tourists in each grid (%) |
| *iCapex$_t$* | Number of year per period *(year)* |
| *LFelect$_p$* | Electrolyzer lifetime *(hours)* |
| *LUT$_l$* | Load and Unload time of product for transportation mode (h/trip) |
| *maxCW* | Maximum water withdrawal rates |
| *minCW* | Minimum water withdrawal rates |
| *n3$_{tm}$* | Discount index applied to capex period |
| *n4$_{t,m}$* | Discount index applied to opex period |
| *Ndm* | Number of days per month |
| *PElec$_{p,i,j}$* | Power capacity of an electrolyzer (kW) |
| *Q$_{ilgg'tm}$* | Quantity of hydrogen transported between grid *g* and grid *g'* |
| *Rsub* | Hydrogen substitution rate for transport application in Corsica (2.5%) |
| *SFCtour$_{tm}$* | Seasonal fuel consumption rate of tourists (%) |
| *SOTWaterVul$_g$* | Vulnerability index for surface groundwater |
| *SP$_l$* | Average speed of transportation mode (km/h) |
| *SRFWaterVul$_g$* | Vulnerability index for surface water |
| *TCAP$_{il}$* | Transport unit capacity (kg/trip) |
| *TMA$_l$* | Availability of transportation mode (h/day) |
| *Tpop$_{gt}$* | Concentration rate of the local population (%) |
| *VulSaison$_{gm}$* | Precipitation risk index |
| *WaterVul$^{max}_{g,m}$* | Maximum water vulnerability index |
| *WaterVul$^{min}_{g,m}$* | Minimum water vulnerability index |

*Continuous variables*

| | |
|---|---|
| *AO$_{gtme}$* | Initial primary energy source availability *(kWh/day)* |
| *ATOT$_{gtm}$* | Primary energy source availability *(kWh/day)* |
| *DRETROFIT$_{igtm}$* | Hydrogen consumption by the retrofitted trucks *(kg/day)* |
| *ESP$_{gme}$* | Energy source power capacity *(kW)* |
| *ESU$_{gtm}$* | Primary energy source consumption rate *(kWh/day)* |
| *IPES$_{egt}$* | Primary energy source importation *(kWh/day)* |
| *IPES$_{egtm}$* | Primary energy source importation *(kWh/day)* |
| *PR$_{p,j,i,g,t,m}$* | Hydrogen production rate *(kg/day)* |
| *PV$_{gme}$* | Availability of photovoltaic energy sources *(kWh)* |
| *WATERCONS$_{g,t,m}$* | Amount of water consumed in each zone *g* *(m$^3$)* |
| *WATERCOST$_t$* | Total daily water cost *(€/day)* |
| *WATERVUL$_g$* | Intermediate water vulnerability index |
| *WATERVUL_SN$_{g,m}$* | Final water vulnerability index |
| *WCV$_{gtm}$* | Total indexed drinking water consumption (Mm$^3$) |
| *WIND$_{gme}$* | Availability of wind energy sources *(kWh)* |

*Integer variables*

| | |
|---|---|
| $IP_{pjigt}$ | New production plants |
| $NP_{p,j,i,g,t}$ | Number of production units |
| $NTUgrid_{ilgg't}$ | Number of transport units |

*Binary variables*

| | |
|---|---|
| $EPSILON_{ilgg't}$ | Binary factor that gives an integer value for the number of transports |
| $XE_{ilg'gt}$ | Electricity flow direction from grid g' to grid g |
| $XE_{ilgg't}$ | Electricity flow direction from grid g to grid g' |

# 1 Introduction

In the backdrop of accelerating climate change, the urgency of the energy transition is paramount, necessitating the adoption of a spectrum of sustainable energy production methodologies. Amidst this evolving technological landscape, hydrogen, derived from renewable sources, such as solar or wind, emerges as a pivotal contender, increasingly garnering attention for shaping the future energy landscape [1]. According to some authors [2], islands can play a significant role in global development by becoming ideal locations to demonstrate new hydrogen pathways for sustainable development, since most islands possess abundant renewable energy resources [3], [4], prompting many to strive for energy self-sufficiency through 100% renewable systems, offering a promising pathway towards carbon neutrality, [5].

Island territories, often distant from continental energy distribution networks, face unique challenges in energy security, environmental sustainability, and high energy supply costs. The use of hydrogen as an energy carrier holds significant potential for these territories, which are currently heavily reliant on fossil fuel imports. Green hydrogen is considered a promising fuel that will significantly contribute to the goals of energy transition, [3]. Hydrogen can limit greenhouse gas (GHG) emissions from conventional transport when used as a clean fuel for fuel cell electric vehicles (FCEVs), which offer longer ranges and faster fueling times than battery electric vehicles (BEVs), [4].

Despite the promise of hydrogen, there is a noticeable gap in the literature regarding the comprehensive deployment of Hydrogen Supply Chains (HSCs), particularly in isolated and non-interconnected areas (NIAs). Existing studies often focus on isolated components of the supply chain or specific case studies, but a holistic approach that considers both the temporal and spatial dynamics of hydrogen production, storage, and distribution remains underexplored. In addition, the process of electrolysis for the production of green hydrogen is heavily dependent on water, a resource that is unevenly distributed across the globe. On average, producing one kilogram of hydrogen requires nearly nine kilograms of demineralized water [6]. A paradox may emerge in regions with potential for renewable energy generation, where the lack of adequate water resources constrains the production of green hydrogen. In this context, local and regional water constraints have been recognized as key factors that define a region or a country's capability to produce and export electrolytic hydrogen. Concentrating hydrogen production in areas subject to water scarcity or to seasonal imbalances can pose additional challenges due to fluctuations in supply and competing demands during dry periods.

This study aims to establish a methodology for designing and operating an HSC system for energy transition in NIAs. The scope of this design study will be comprehensive, encompassing the entire geographic territory of the island to determine an optimal value chain for hydrogen production, while identifying all necessary infrastructure and their strategic locations throughout the territory.

This includes considering seasonal variations in certain parameters, primarily water and energy resources.

This study builds upon our previous work [7] by introducing several critical innovations. First, the current work explicitly integrates potable water availability into the modeling framework, responding to the resource constraints that are particularly relevant in island territories. Second, it departs from previous assumptions by relying exclusively on renewable energy sources, fully decoupled from the electrical grid. Third, hydrogen demand is modeled dynamically based on tourism activity, thus introducing inter-seasonality into both production and distribution planning. Finally, the optimization strategy is extended from a single-objective to a multi-objective framework, incorporating cost, environmental impact, and risk criteria, and using the M-TOPSIS method for robust decision support. These contributions aim to improve the realism and transferability of the proposed model for sustainable hydrogen deployment in resource-limited regions.

The key research questions addressed in this study are the following:
1. How can the HSC be optimally designed and managed to account for seasonal and spatial variations in renewable energy and water resources?
2. What is the optimal configuration of production, storage, transport, and distribution facilities to ensure economic viability and sustainability?
3. How can water resource management be integrated into the HSC framework to address the specific challenges of island territories?
4. What are the impacts of seasonal tourist activities on hydrogen demand and how can these be factored into the supply chain design?
5. How can the developed framework be adapted and applied to other isolated territories to support broader energy transition goals?

For this purpose, a methodological framework is developed to simultaneously address the optimal design and management of a green HSC by focusing on several aspects:
- **Supply Chain Design and Management:** The problem is based on MILP formulation using the minimization of three objective functions considered in isolation and simultaneously: total system cost, GHG emissions, and an index based on hydrogen risk. Particular attention is focused on the temporal resolution adapted to the modeling objective, as consumption patterns vary greatly depending on the season.
- **Spatial Resolution:** The spatial resolution is addressed by coupling both upstream and downstream of the MILP with a Geographic Information System (GIS) approach, respectively, to identify the hydrogen demand area and the potential production, storage, and distribution sites, as well as to validate the optimal solutions obtained and position the different units within the territory.
- **Water and Energy Resource Assessment:** This study will also consider the assessment of surface water and groundwater availability in each region and for each month of the year, as well as the needs for local energy resources.
- **Reduced GHG emissions:** The use of retrofitted hybrid vehicles, which are hydrogen-powered electric vehicles for hydrogen transport, will also be studied to further reduce GHG emissions.

- **Multi-objective optimization:** The optimization strategy will also be extended to a multi-objective approach using an ε-constraint method, followed by the use of a multi-criteria decision support tool based on the m-TOPSIS method.

This paper is organized into five sections. Section 1 provides a literature review of research studies that have examined the development of HSC, particularly in isolated territories and water resource model optimization. Section 2 outlines the general methodological framework of the study and the developed mathematical model. Section 3 presents the case study of Corsica, including the various specific parameters of the territory. Section 4 presents the different optimization results obtained. Section 5 draws conclusions and perspectives.

## 2 Literature review

### 2.1 HSC Modeling and optimization approaches

HSCs involve multiple stages, including production, storage, distribution, and end-use applications [8]. While many studies optimize these stages individually, integrated and sustainable planning remains challenging, particularly in complex environments. Optimization methodologies, such as the MILP approach [9], [10], stochastic modeling [11], [12], [13], [14] and multi-objective approaches [11], have been used to improve techno-economic feasibility. However, these approaches are often limited to cost minimization or bi-criteria optimization (e.g., cost-risk, cost-environment) and rarely incorporate dynamic or geographical factors in depth.

Recent works have introduced finer temporal resolutions and spatial constraints to address seasonal demand variability and local limitations [15]. Some studies include rolling horizon methods to capture investment decisions across a decade, while others factor in urban infrastructure, natural site constraints, and traffic data [12], [13], [14]. Still, the operational feasibility of hydrogen supply systems under complex real-world constraints remains underexplored.

Recent studies have also emphasized the role of multi-criteria decision analysis (MCDA) in the planning of green hydrogen supply chains. For instance, [16], [17] demonstrate how combining technical, economic, and environmental criteria enhances the strategic selection of renewable energy sites for hydrogen production. These approaches provide valuable insights into spatial prioritization, especially for regions facing resource limitations or policy constraints. However, most of these MCDA-based methods still lack integration into fully optimized, inter-seasonal, and geographically explicit models, which our study seeks to address.

### 2.2 Hydrogen system deployment in NIAs

Island regions pose specific challenges to hydrogen deployment due to geographic isolation, limited energy infrastructure, and resource constraints. Studies such as Dimou et al. [18] examined the case of Anafi Island in Greece, demonstrating that with an approximate 84% penetration rate of renewable energies, 40% of the total renewable energy produced would be surplus. This excess energy could then be used to produce hydrogen for the public transport fleet (4 buses) and around 200 additional light vehicles. Carere et al. [19] investigated a 100% renewable energy solution for Sardinia, while Meza et al. [20] proposed a combination of photovoltaic solar panels and wind turbines for transitioning a small Nicaraguan island to 100% green energy. Berna-Escriche et al. [21] analyzed the feasibility of meeting 100% of the projected energy demand for 2040 in the Canary Islands using photovoltaic solar panels, wind turbines, pumped storage, and battery storage.

The surplus energy from these sources is utilized to produce hydrogen to satisfy non-electric demands such as transport and industry.

Seasonal variations significantly impact the feasibility of renewable hydrogen production, particularly in island environments. Corsica, for example, experiences fluctuations in solar and wind energy generation, affecting the consistency of hydrogen production. Studies such as in [22] have examined seasonal storage solutions to mitigate these fluctuations, including hydrogen storage in underground caverns. Additionally, geographic factors, including the availability of coastal versus inland water sources, influence the feasibility of electrolysis-based hydrogen production. While previous research has addressed seasonal energy storage, the specific challenges posed by inter-seasonal water availability for hydrogen production remain underexplored.

### 2.3 Water-Energy nexus and the role of resource constraints

Hydrogen production through electrolysis demands substantial water inputs, raising sustainability concerns in water-scarce regions. As highlighted in [23], [24], water availability remains a critical but often overlooked constraint in HSC modeling. In island territories, where freshwater is limited and often sourced through desalination, this issue becomes particularly acute. Some studies have explored coupling electrolysis with desalination systems [25], but few address the full economic and environmental trade-offs in such remote contexts.

Furthermore, regulatory and policy frameworks like the EU Hydrogen Strategy [26], [27] promote sustainable hydrogen development but offer limited guidance on water governance for electrolysis projects. In regions like Corsica, local water management policies could significantly influence the viability of large-scale hydrogen production. While some global analyses suggest that water demand for hydrogen is small compared to food production ($<1.5\%$), this generalization overlooks critical local constraints on freshwater access and distribution.

To our knowledge, the literature lacks a comprehensive HSC model integrating both energy and water resource planning for isolated territories. This study addresses this gap by developing a multi-objective, inter-seasonal optimization framework that explicitly considers potable water availability in the deployment of hydrogen infrastructure in Corsica. To clearly highlight the existing research gaps and contextualize our contribution, Table 1 summarizes the key features of representative studies on HSC modeling, with a focus on approaches, objectives, and the integration of local constraints.

| Study (Authors) | Territory / Context | Modeling type | Objective function(s) | Local constraints considered |
|---|---|---|---|---|
| **Ming et al. [8]** | Dalian city in China | MILP | Cost and GHG minimization | Demand variability, Energy access |
| **Dimou et al. [18]** | Anafi Island (Greece) | MILP | Cost minimization | Energy access |
| **Carere et al. [19]** | Sardinia | MatDyn | Frequency control, 100% Renewable Scenario | Energy infrastructure Inertial dynamics |
| **Meza et al. [20]** | Ometepe Island (Nicaragua) | Simulation-based | 100% Renewable Scenario | Energy access |
| **Berna-Escriche et al. [21]** | Canary Islands | HOMER | 100% Renewable Scenario | Energy access, Demand variability |
| **Kim and Moon [11]** | Korea | Multi-objective MILP | Cost vs. Safety | Optimal infrastructure, Demand uncertainty |
| **Mao et al. [28]** | China | MILP + stochastic | Cost minimization | Optimal infrastructure |
| **Camelo et al.[29]** | Region/Bresil | MILP | Cost minimization | Optimal infrastructure |
| **Goh et al. [30]** | Peninsular Malaysia | MILP | Cost, GHG emissions, Risk | Demand variability, temporal resolution |
| **Current Study (This paper)** | Corsica (France – Island) | Multi-objective MILP | Cost, GHG emissions, Risk, Water constraints | Water availability, land use, Energy access |

*Table 1 Summary of previous work vs this study*

## 3 Method and tools
### 3.1 General framework

Figure 1 illustrates the proposed methodology for the optimal design and operation of hydrogen energy systems, taking into account both spatial and temporal resolutions through a multi-objective optimization framework. The framework of the investigated energy systems is defined by determining renewable power profiles, estimating hydrogen demands, and gathering technical, economic, environmental and, if required, geo-referenced data for the potential technologies. For this purpose, a geospatial approach using the QGis tool [14] has been developed. It enables the identification of suitable sites for hydrogen infrastructure such as production, storage, and distribution facilities along with associated transport networks, based on geographical constraints as well as the current location of strategic power plants and policies for the implementation of new ones. A set of techno-economic parameters is considered: production, storage, and transport options, possible locations, available energy sources, capital and operation cost, technical features (efficiency, capacity, lifetime, load factors, storage capacities), GHG emissions and relative risk index for the various technologies. Next, scenarios for hydrogen deployment options in the mobility sector are studied.

*Temporal discretization:* Typically, long-term energy system models do not simulate every hour or day of the year in detail, but instead use a limited set of representative time slices. Each time slice reflects characteristic demand and supply patterns over part of the year. This coarse temporal resolution is generally acceptable for long-term studies and helps maintain reasonable computational performance. The detailed assessment of the daily distribution could create a false precision with respect to the overall uncertainties associated with long-term projections [31]. In this work, the time horizon has been divided into months to capture the evolution of demand from 2025 to 2050, while investments and operating costs are accounted for at the beginning and end of each year, respectively, according to economic analysis rules.

*Spatial discretization:* the spatial approach is most often conducted by dividing the territory under consideration according to its administrative distribution. The distribution can be based on the number of municipalities or regions depending on the size of the territory and the detail of the mesh we need. The finer the mesh, the more precise the results and the longer the simulation time. A distribution according to the clusters of municipalities, formed to address among others common problems in the field of energy, water and the environment by adopting an appropriate economic and organisational policy. There are 9 local authorities in Corsica where the number, size, and type of production and storage units (integer variables) must be determined according to the considered objective functions (total annual cost, GHG emissions, and risk index) [32] and constraints (e.g., mass balances, demand satisfaction) as well as the flow rate (continuous variables) of hydrogen produced, stored and transported into the network, [33].

The optimization problem for the design and operation of the HSC developed in this work is formulated as a MILP model implemented in the GAMS® modeling system, solved using CPLEX 12 [32].

To incorporate the multi-criteria nature of the problem, the ε-constraint method was employed, considering three objectives simultaneously to generate a three-dimensional Pareto front. In this formulation, the minimization of the cost objective is treated as the primary objective function, while environmental impact and safety are introduced as inequality constraints. The optimal trade-off solution among the multi-objective set was identified by calculating the shortest distance from the Utopia point to all Pareto-optimal solutions using the TOPSIS method [32]. Subsequently, a post-optimization analysis is performed on the selected compromise solutions to verify their technical feasibility using the GIS tool.

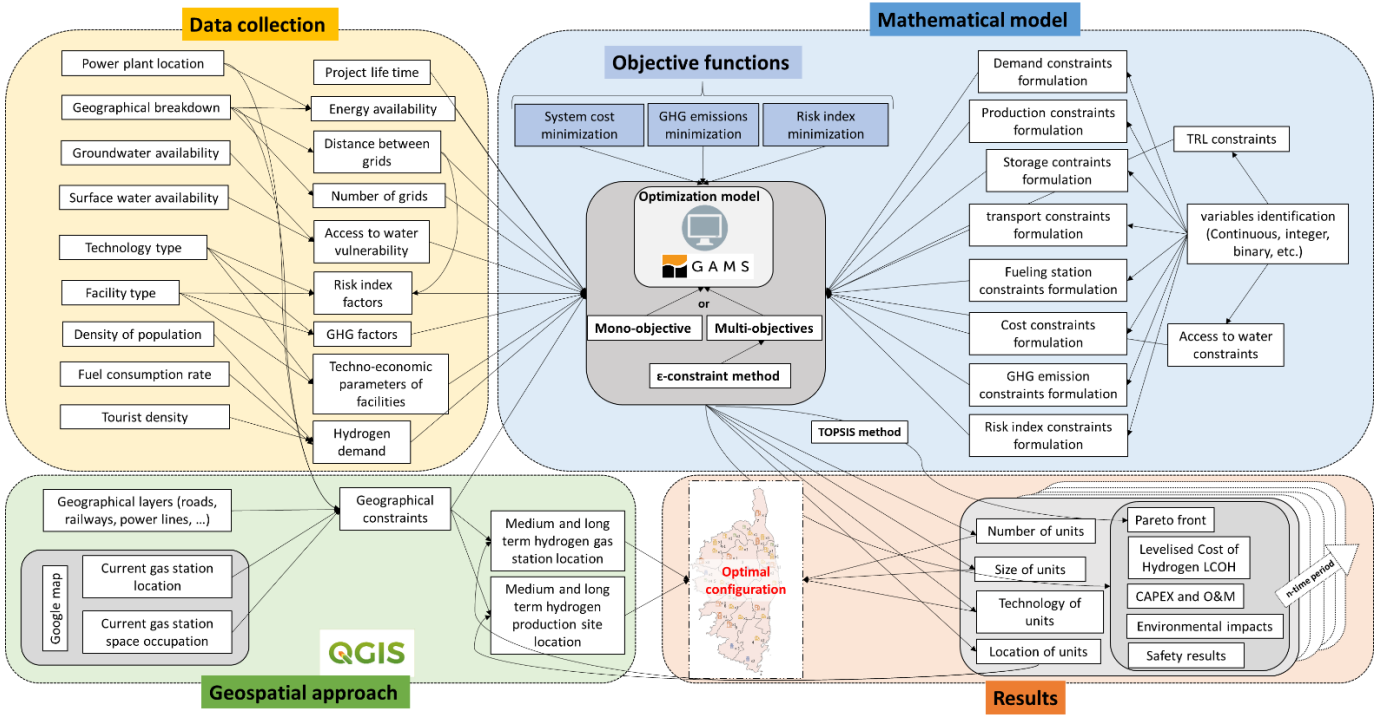

*Figure 1: Methodological framework overview*

## 3.2 Mathematical constraints

The general mathematical formulation of this model is similar to that proposed in the previous mathematical model. Some constraints of this model refer to those of the previous publication with the integration of monthly temporality [7]. These include hydrogen demand constraints, storage constraints, transport and distribution constraints, as well as GHG emissions and risk index objective functions. This paper will therefore only present the constraints specific to the current model. All new or modified constraints are indicated by an (*).

### 3.2.1 Energy supply constraints.

The total availability of primary energy sources in a given zone (or grid) *g* during period *t* et and for month *m* ($ATOT_{gtm}$) is defined as the sum of three components: the initial average availability of renewable energy sources in each zone and for each month $A0_{gtme}$, the importation of primary energy sources from the grid $IPES_{egtm}$ and the consumption rate of primary energy sources $ESU_{gtm}$.

$$ATOT_{gtm} = \sum_e (A0_{gtme} + IPES_{egtm}) - er \times ESU_{gtm} \;; IPES_{egtm} = 0; \forall\, e,t,g,m \;; g \neq g'$$

*Equation 1*

$ESU_{gtm}$ is defined as the product of $\gamma_{pj}$ which represents the energy consumption rate of production facities, and $PR_{pjigtm}$, the daily hydrogen production rate for each type of production facility *p* and size *j*.

$$ESU_{gtm} = \sum_{pji}(\gamma_{pj} \times PR_{pjigtm}) \ \forall \ g,t,m \ ; \quad g \neq g' \qquad \text{Equation 2}$$

The total availability of photovoltaic and wind energy sources ($PV_{g,m,e}$ et $WIND_{g,m,e}$) is determined by considering the power capacity of the energy source in each network during period $t$ and for each energy type ($ESP_{gte}$), the number of hours per month $Mh_m$, the capacity factor for each energy source ($PVCF_{et}$ et $WindCF_{et}$, respectively for photovoltaics and wind energy), and the total number of days per month $Nd_m$.

$$PV_{g,m,e} = \frac{ESP_{g,m,e} \times Mh_m \times PVCF_m}{Nd_m} \qquad \forall \ g,m,e; \quad g \neq g' \qquad \text{*Equation 3}$$

$$WIND_{g,m,e} = \frac{ESP_{g,m,e} \times Mh_m \times WindCF_m}{Nd_m} \qquad \forall \ g,m,e; \quad g \neq g' \qquad \text{*Equation 4}$$

$AO_{g,t,m,e}$ represents the total availability of renewable energy sources and is defined as the sum of the availability of renewable energy sources ($PV_{g,m,e}$ et $WIND_{g,m,e}$) multiplied by the capacity evolution rate of these energy sources during period $t$ and for each renewable energy category, $eCo_{te}$.

$$AO_{g,t,m,e} = (PV_{g,m,e} + WIND_{g,m,e}) * eCo_{t,e} \quad \forall \ g,t,m,e; \quad g \neq g' \qquad \text{*Equation 5}$$

### 3.2.2 Production constraints

The minimum and maximum daily production capacities ($PCAP^{min}_{p,I,j}$ et $PCAP^{max}_{p,i,j}$) are calculated based on the lower and upper daily capacity factors of the electrolyzer of type $p$ and size $j$, expressed in hours per month ($ELCF^{min}_{p,j,i}$ et $ELCF^{max}_{p,j,i}$, respectively) and the assigned power of each electrolyzer, given in kW, $pElec_{p,i,j}$. $Nd_m$ represents the number of days per month and $\gamma_{pj}$ represents the energy consumption rate of production facilities (kWh/kg $H_2$).

$$PCap^{min}_{pij} = \frac{\left(\frac{ELCF^{min}_{pij}}{Nd_m}\right) \times PElec_{pij}}{\gamma_{pj}} \ ; \ \forall \ p,i,j \qquad \text{*Equation 6}$$

$$PCap^{max}_{pij} = \frac{\left(\frac{ELCF^{max}_{pij}}{Nd_m}\right) \times PElec_{pij}}{\gamma_{pj}} \ ; \ \forall \ p,i,j \qquad \text{*Equation 7}$$

### 3.2.3 Geographical constraints

Geographical constraints for the location of units have been integrated into the model based on a GIS-based study, specifically using QGIS.
The installation of new production plants $IP_{p,j,i,g,t}$ in a given zone $g$ can only be implemented from a specific period onward, depending on the projected locations identified through the GIS-based method. For instance, some grids cannot accommodate new production systems before 2030.

$$IP_{pjigt} = 0, \quad \forall p,j,i,g = \{4; 5\} \ et \ t = \{1; 2; 3; 4; 5\}; \qquad \text{*Equation 8}$$

In addition, some grids are designated for photovoltaic (**PV**) power plants, while others grids can accommodate wind farms (**WIND**). All other zones are restricted from hosting either PV power plants or wind farms.

$$ESP_{gme} = 0, \forall m, g = \{1; 2; 3; 4; 5; 8; 9\}, e = \{PV\} \qquad \text{*Equation 9}$$

$$ESP_{gme} = 0, \forall m, g = \{2; 3; 5; 6; 8; 9\}, e = \{WIND\} \qquad \text{*Equation 10}$$

Medium and large-scale production units can only be installed in zones where renewable energy sources are available, as they are considered centralized production units. In all other zones, only small- and mini-scale units can be deployed.

$$IP_{pjigt} = 0, \quad \forall p, i, t, j = \{Large; med\}; \ g = \{2; 3; 5; 8; 9\}; \qquad \text{*Equation 11}$$

Hydrogen export can only occur in one direction, from centralized to decentralized production centers (small and mini-scale units). ***NTUgridm$_{i,l,g,g',t}$*** is the number of transport unit.

$$NTUGRID_{i,l,g,g',t,m} = 0, \quad \forall i, l, g, j = \{Large; med\}; \ g = \{2; 3; 5; 8; 9\}; \qquad \text{*Equation 12}$$

### 3.2.4 Constraints related to the "retrofit"

In this article, an alternative transportation mode is introduced, based on retrofitted electric trucks powered by hydrogen. The concept involves repurposing used heavy-duty vehicles by replacing the internal combustion engine with an electric motor. The electric motor will be powered by a fuel cell, with electrochemical battery storage and compressed hydrogen (at high pressure) used to store energy, thereby increasing the vehicle's range.

This system reduces investment costs associated with transportation while also lowering GHG emissions during the usage phase. According to some manufacturers, the complete retrofitting system (vehicle + equipment) costs approximately 30% to 40% less than a new vehicle of the same model [34]. The characteristics of the retrofitted vehicles are provided in appendix B.

When considering this scenario, all transportation means essential for the operation of the HSC are assumed to be replaced by retrofitted vehicles.

The hydrogen consumption by the retrofitted trucks ***DRETROFIT$_{i,g,t,m}$*** is calculated by multiplying the number of transport units established between grid ***g*** and grid ***g'*** ***NTUGRID$_{i,l,g,g',t,m}$*** by the average distance between grids ***AD$_{g,g'}$*** and the hydrogen consumption rate of the vehicle ***FCEVcons***. Equation 13 presents this constraint:

$$DRETROFIT_{i,g,t,m} = \sum_{l,g'} \frac{NTUGRID_{i,l,g,g',t,m} \times 2 \times AD_{g,g'} \times FCEVCons}{100} \ ; \forall i, g, t, m \ ; \qquad \text{*Equation 13}$$

The number of transport units ***NTUGRID$_{i,l,g,g',t,m}$*** is obtained from the quantity of hydrogen transported between grid ***g*** and grid ***g'***, ***Q$_{i,l,g,g',t}$***.

$$NTUGRID_{i,l,g,g',t,m} = \frac{Q_{i,l,g,g',t,m}}{TMA_l \times TCAP_{i,l}} \times \frac{2 \times AD_{g,g'}}{SP_l + LUT_l} + EPSILON_{i,l,g,g',t} \ ; \forall i, l, g, g', t \ ; \qquad \text{*Equation 14}$$

### 3.2.5 Water resource constraints

Water availability is a critical factor in hydrogen production through electrolysis, as approximately 9 liters of water are required to produce 1 kg of hydrogen [6]. In regions with limited water resources, competition between hydrogen production and other water uses (domestic, agricultural, and industrial) must be carefully managed. The Mediterranean region, for example, experiences significant water stress, which is further intensified by tourism, leading to strong seasonal demand fluctuations. Studies highlight the impact of tourism on water availability and its competition with local users [30] [31].

To integrate water resource constraints into HSC modeling, it is essential to assess both surface and groundwater availability, as well as seasonal variations in water resources. This assessment allows the formulation of constraints that prioritize hydrogen production in zones and periods with lower vulnerability to water shortages.
These constraints related to water resources are specific to the studied territory, in this case, Corsica. They are directly linked to the geographical and climatic specificities of the region. The constraints identify all water-stressed areas within the territory using a global water vulnerability index, which will be presented in the case study.

### 3.2.6 Cost objective function

The investment costs (CAPEX) for the subsystems of production, storage, transport, and distribution remain identical than the equations described in [7] since the estimation of these costs is determined based on annual evaluations $t$ (no variable depends on the index $m$).
The maintenance and operational costs (OPEX) for this model are estimated for each monthly period $m$. The constraints used take into account the monthly hydrogen production rate $PR_{p,j,i,g,t,m}$, storage $ST_{i,g,t,m}$, distribution $FR_{fs,j,i,g,t,m}$, transport $Q_{i,l,g,g',t,m}$, electricity and water consumption rates ($IPES_{e,g,t,m}$, $RESU_{e,g,t,m}$ and $WATERCOST_t$).

$$LCOH = \sum_{m,t} \frac{\left[\frac{CAPEX_t}{[(1+f)\times(1+dr)]^{n3_{t,m}}}\right] + \left[\frac{OPEX_t \times Nd_m}{[(1+f)\times(1+dr)]^{n4_{t,m}}}\right]}{\frac{DTtot_t}{[(1+f)\times(1+dr)]^{n4_{t,m}}}}; \forall t \qquad *Equation\ 15$$

The time progression index ($n3_{t,m}$) evolves in this model according to the year and the month in question. To update the first part of the Levelized Cost of Hydrogen (LCOH) equation (the CAPEX part), the index ($n3_{t,m}$) is only available for the month of January of each study year and is zero for all other months. Investments can only occur at the beginning of the year.
Regarding the update of variable costs (OPEX), the index ($n4_{t,m}$) remains constant for a given year and changes when moving to the next year. This ensures that the same discount rate is applied to all monthly expenses within the same year.

### 3.2.7 GHG emission objective function.

Total daily GHG emissions **GWPtot**, which represent the amount of emissions throughout the logistics chain, are obtained by summing the emissions produced from production **PGWP$_t$**, storage **PGWS$_t$**, and transportation **PGWT$_t$** sectors.

$$GWPtot = \sum_t PGWP_t + SGWS_t + TGWT_t \qquad Equation\ 16$$

### 3.2.8 Risk index objective function.

The total risk of the hydrogen logistics chain, denoted **RiskTotal**, is obtained by summing the three risk indices corresponding to the three sectors: production **TPRisk$_t$**, storage **TSRisk$_t$**, and transport **TTRisk$_t$**.

$$RiskTotal = \sum_t TPRisk_t + TSRisk_t + TTRisk_t \qquad \text{Equation 17}$$

## 4 Supply chain model for the Corsica case study

### 4.1 Problem description and assumptions

The supply chain model developed for Corsica is based on a set of simplifying assumptions, illustrated in Figure 2, and reflects the territorial characteristics and infrastructure constraints.

- The island is divided into 9 spatial grids, denoted g when acting as a hydrogen exporter and g′ when acting as an importer. Each grid represents a distinct economic, social, and physical area where hydrogen related projects may be deployed. Each grid is characterized by a hydrogen demand profile and an energy production potential.
- The model considers a long-term planning horizon up to 2050, with monthly resolution to capture the temporal fluctuations in hydrogen production and delivery across the territory.
- Hydrogen production units can be installed only in selected grids (identified by a prior GIS-based suitability analysis). Two electrolyser technologies are considered: Proton Exchange Membrane (PEM) and Alkaline (AE).
- Hydrogen storage units may be installed in all grids, and are available in four capacity classes (mini, small, medium, large), consistent across production and storage technologies.
- Each grid is assumed to have 3 days of hydrogen storage autonomy to buffer local fluctuations.
- Hydrogen can be:
  - Compressed and transported via tanker trucks to other grids (g′)
  - Liquefied and transported via tube trailers to distant grids (g′)
- Hydrogen distribution units can be installed in any grid that has a non-zero demand.
- Only photovoltaic (PV) and wind energy are considered as sources of electricity input for hydrogen production.
- No additional investment for the electrical grid is required as long as the total installed capacity of electrolyzers does not exceed 200 MW.

Each locality is assigned a specific hydrogen demand distributed over all time periods from 2026 to 2050. Hydrogen is produced using the available renewable electricity and water within a given grid g, at a pressure of 35 bar, and stored in low-pressure tanks (30-35 bar) as a backup in case of production failure. When local hydrogen production exceeds the demand in a given zone, the surplus is compressed and transported by truck to another grid g′ experiencing a production shortfall.

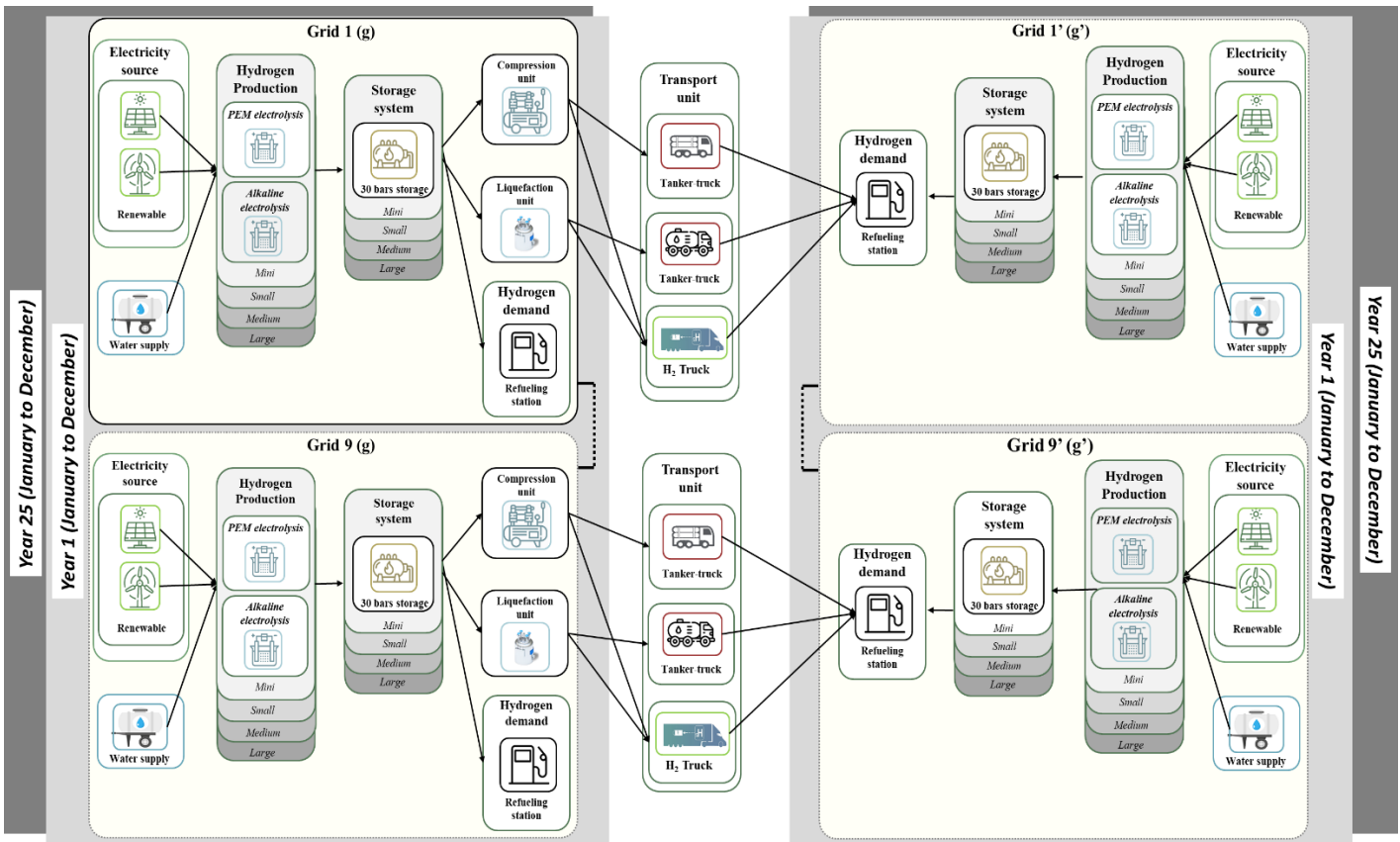

*Figure 2: Corsica supply chain superstructure scenario*

The model determines, for each period:
- The number, technology, size and location for the production, storage and distribution units;
- The amount of hydrogen produced, stored and distributed;
- The amount of hydrogen transported and the importing or exporting grid;
- The operating hours of each production unit;
- The capital and operating cost of all facilities and transport units;
- The amount of GHG emissions for the facilities and transport units;
- The risk level index of the facilities and transport units;
- The average cost and LCOH over the life time of the project.

These decisions are made while optimizing one or more of the following objective functions:
- Minimization of total cost, including both capital and operating expenditures across the HSC;
- Minimization of total GHG emissions, accounting for all stages of hydrogen production, storage, and transport;
- Minimization of the total risk index, aggregating the risk contributions from each facility and

To address the multi-objective nature of the problem, an ε-constraint method is applied, followed by a multi-criteria decision support analysis using the modified TOPSIS (m-TOPSIS) approach. Equal weighting coefficients are assigned to the three criteria: cost, GHG emissions, and risk index.

## 4.2 Data collection

### 4.2.1 Hydrogen Demand.

The hydrogen demand is estimated based on the current fuel consumption in the transport sector, which includes private passenger vehicles, public transport, freight transportation, and transport associated with tourism-related activities. Figure 3 illustrates the relationship between these parameters and their contribution to the overall hydrogen demand.

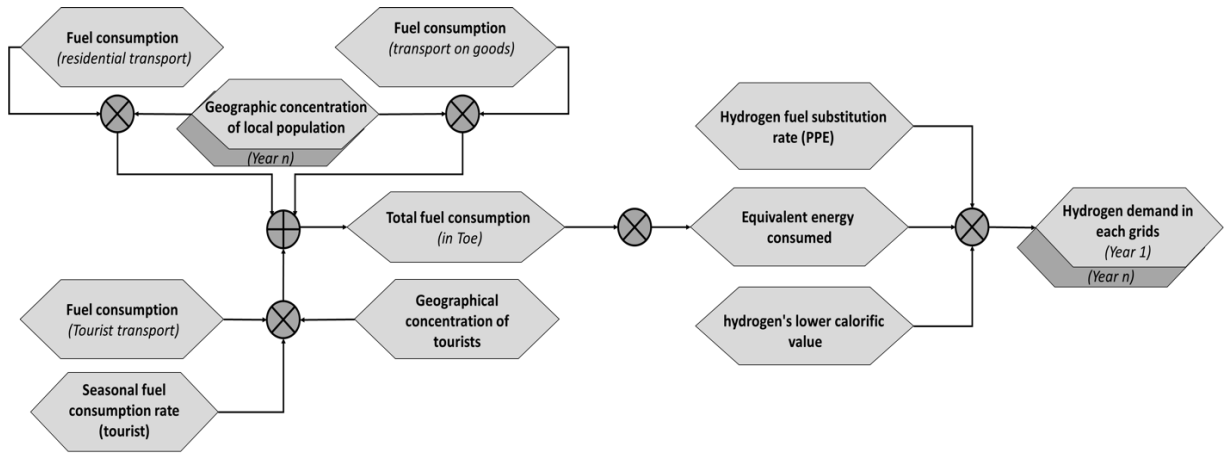

*Figure 3: Hydrogen demand evaluation*

According to the Pluriannual Energy Programme (PPE), Corsican authorities plan to substitute 2.5% of the total fuel consumption by hydrogen to support the deployment of FCEV fleet in 2030 [35]. Beyond 2030, hydrogen demand is projected to evolve according to population growth, following demographic trends published by the French National Institute of Statistics (*INSEE* [36]).

The influence of tourism is explicitly integrated into the hydrogen demand evaluation. It is based on detailed data regarding the number of tourist accommodations (beds per type of lodging) and overnight stays per accommodation type. This dataset, collected at the municipal level by the local tourist agency, was spatially aggregated and redistributed over the 9 model grids using *QGIS*. It is important to note that the most densely populated grids are not necessarily the most visited by tourists: as shown in Figure 4, grids 6 and 7 exhibit the highest resident populations, while grids 1 and 4 register the highest tourist activity.

The blue curve in Figure 4 represents the final hydrogen demand profile, which combines both residential and tourism-related mobility demand. This profile is used as input in the model and is represented by a colour intensity gradient. It captures the seasonal variations in hydrogen demand driven by tourism. Notably, approximately 20% of the island's total fuel consumption is attributable to tourism-related transport [37], primarily concentrated in the summer months, thus emphasizing the need for a dynamic and spatially resolved demand model. Further details on tourist distribution by grid are provided in Appendix B.

Equation 16 expresses the hydrogen demand for a given grid *g*, over a the time period *t* end *m*, denoted as $DH2_{g,t,m}$. It is calculated based on fuel consumption related to the transportation of residents, goods, and tourists, represented respectively by **FCres**, **FCgds** et **FCtour**.

The parameter $Tpop_{g,t}$ represents the concentration rate of the local population in grid *g* for period *t*. Similarly, $SFCtour_{t,m}$ and $Gtour_g$ correspond respectively to the seasonal fuel consumption rate of tourists, and the concentration rate of tourists in each grid.

Finally, **Etoe** represents the conversion factor of one ton of oil equivalent (toe) into kilowatt-hours, with 1 toe equating to 11,630 kWh, while **Rsub** denotes the hydrogen substitution rate, set at 2.5%.

$$DH2_{g,t,m} = [(FCres + FCgds) \times Tpop_{g,t} + (FCtour \times SFCtour_{t,m} \times Gtour_g)] \times 1000 \times Etoe \times PCI \times Rsub \ ; \ \forall \ g,t,m \qquad *Equation\ 18$$

Table 2 presents the average hydrogen demand required for each period and each grid. Grids 6 and 7 exhibit the highest levels of demand, primarily due to their high population density and associated transport needs. Interestingly, these zones do not coincide with the areas showing the highest tourist activity, confirming that resident mobility remains the dominant factor in hydrogen demand for these grids.

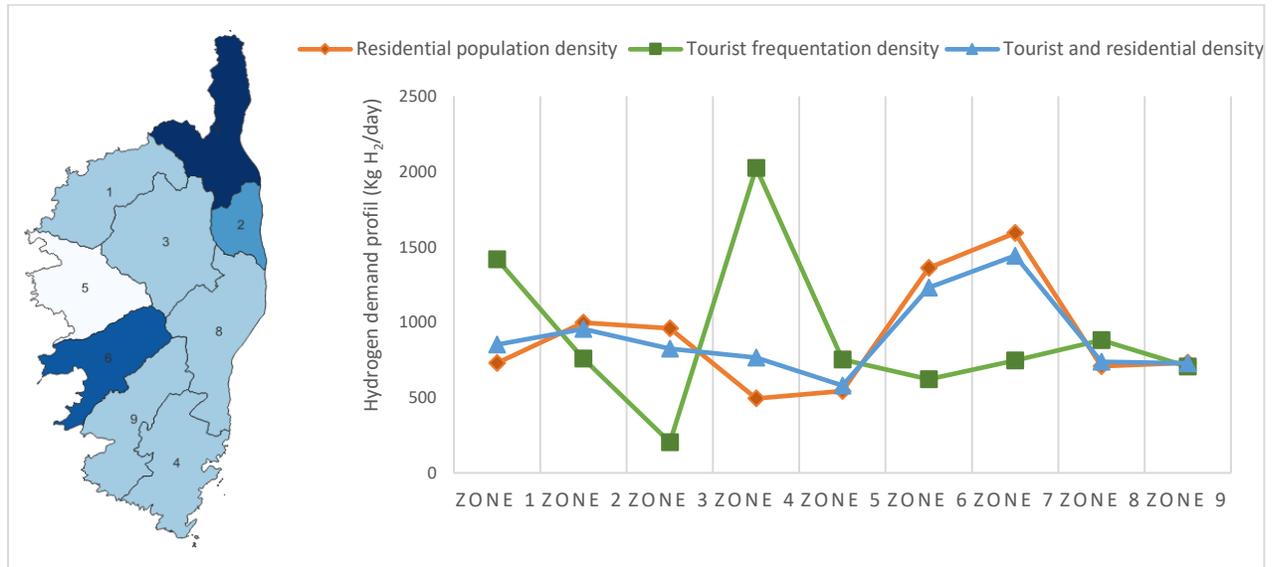

*Figure 4: Hydrogen demand grid distribution considering different profiles*

| $DH2_{g,t,m}$ | Grid 1 | Grid 2 | Grid 3 | Grid 4 | Grid 5 | Grid 6 | Grid 7 | Grid 8 | Grid 9 | Total |
|---|---|---|---|---|---|---|---|---|---|---|
| 2025 (kg/day) | 382 | 429 | 370 | 344 | 260 | 552 | 647 | 332 | 327 | 3,643 |
| 2030 (kg/day) | 558 | 626 | 541 | 502 | 380 | 806 | 945 | 484 | 477 | 5,319 |
| 2035 (kg/day) | 854 | 958 | 827 | 768 | 582 | 1,233 | 1,446 | 740 | 729 | 8,137 |
| 2040 (kg/day) | 1,282 | 1,439 | 1,242 | 1,153 | 873 | 1,852 | 2,171 | 1,112 | 1,095 | 12,219 |
| 2045 (kg/day) | 1,925 | 2,161 | 1,866 | 1,732 | 1,312 | 2,781 | 3,261 | 1,670 | 1,645 | 18,353 |
| 2050 (kg/day) | 2,892 | 3,246 | 2,803 | 2,601 | 1,970 | 4,178 | 4,899 | 2,509 | 2,470 | 27,568 |

*Table 2: Average hydrogen demand per grid*

### 4.2.2 Water availability and constraints

The Mediterranean is the most visited tourist region in the world, with nearly 300 million arrivals per year [38]. It has already been identified as an area where tourism places considerable pressure on water supply, competing with local users [39]. The Mediterranean is also the most studied region in terms of water use in tourism [40].

To incorporate water availability constraints into the model, we assessed resource availability using two key indicators: surface water and groundwater availability. These data are derived from the study conducted by the Corsican Basin Committee [41].

Overall, it can be observed that the northern part of the island, represented by zones 1, 2, 3, 7, and the extreme south (zone 4), are the areas where both surface and groundwater resources are the most vulnerable. In the western part of the island, zones 5 and 6 are notably vulnerable in terms of groundwater availability but normal with respect to surface water.

On a scale of 1 to 3, representing the level of vulnerability of each zone, similar to the Corsican Basin Committee study [41], the final vulnerability of each zone is obtained by multiplying the vulnerability level associated with surface or groundwater resources by the consumption rate of these waters. According to the Corsican Hydraulic Equipment Office (OEHC), 80% of the water used in Corsica is surface water, and 20% is groundwater [42]. Table 3 shows the degree of vulnerability of the different zones based on these parameters. Level 1 corresponds to low vulnerability, while level 3 represents high vulnerability.

Equations 17 and 18 are used to determine the intermediate vulnerability value $WATERVUL_g$ and the final vulnerability value $WATERVUL\_SN_{g,m}$ of each grid. $SRFWaterVul_g$ and $SOTWaterVul_g$ represent the vulnerability index for surface water and groundwater, respectively.

$$WATERVUL_g = 0{,}8 \times SRFWaterVul_g + 0{,}2 \times SOTWaterVul_g; \ \forall \ g \qquad \text{*Equation 19}$$

According to Météo France, Corsica receives approximately 8 billion m³ of water annually in the form of precipitation, with 86 million m³ consumed (1.1%) [42]. These precipitations are unevenly distributed both temporally (low during the summer) and geographically (low along the coast). In addition to these constraints, tourist pressure increases during the same periods and is primarily concentrated along the coast. In total, five (5) areas are affected by this influx, notably zones 1, 3, 4, 6, and 7 (see table 2). For each of these zones, the hypothesis made is to add an additional constraint indicating the level of pressure on water availability during certain times of the year. Using data from Météo France, we selected the months of the year when the precipitation risk is below 40%, specifically May, June, July, August, and September ($VulSaison_{gm}$). Zones with a precipitation shortage index of 2 correspond to areas where the vulnerability of surface and groundwater accessibility is doubled.

$$WATERVUL\_SN_{g,m} = WATERVUL_g \times VulSaison_{g,m} \qquad \text{*Equation 20}$$

| Grids | Precipitation deficit index ($VulSaison_{g,m}$) | | Surface water vulnerability index ($SRFWaterVul_g$) | groundwater vulnerability index ($SOTWaterVul_g$) | final water vulnerability index (*$WATERVUL\_SN_{g,m}$) | |
|---|---|---|---|---|---|---|
| | Summer | Winter | | | Summer | Winter |
| 1 | 2 | 1 | 3 | 2 | 5,6 | 2,8 |
| 2 | 1 | 1 | 2 | 3 | 2,2 | 2,2 |
| 3 | 2 | 1 | 2 | 2 | 4 | 2 |
| 4 | 2 | 1 | 3 | 3 | 6 | 3 |
| 5 | 1 | 1 | 1 | 3 | 1,4 | 1,4 |
| 6 | 2 | 1 | 1 | 2 | 2,4 | 1,2 |
| 7 | 2 | 1 | 3 | 3 | 6 | 3 |
| 8 | 1 | 1 | 2 | 1 | 1,8 | 1,8 |
| 9 | 1 | 1 | 1 | 1 | 1 | 1 |

*Table 3: Water vulnerability index for each grid and season*

To constrain the model to produce hydrogen in zones where the vulnerability level is low, and complementarily in zones and periods where water availability is greatest, lower and upper limits for the water vulnerability index ($WaterVul^{min}_{g,m}$; $WaterVul^{max}_{g,m}$) and water withdrawal rates ($minCW$ et $maxCW$) have been set. These limits apply to the product of the amount of water consumed in each zone $g$ and for each period $t$ and $m$, $WATERCONS_{g,t,m}$, and the final vulnerability level of each zone involved for each period, $WATERVUL\_SN_{g,m}$. The minimum and maximum final water vulnerability index ($WATERVUL\_SN_{g,m}$) are respectively set at 1 and 5. When considering a maximum index value of 5, water withdrawal restrictions will only apply when both the precipitation shortage index and the water accessibility vulnerability index (for both surface and groundwater) are simultaneously at their maximum.

If the maximum final vulnerability index is set to 4, it means that no restrictions will be applied unless the precipitation shortage index is at its maximum ($VulSaison_{g,m}=2$) and at least one of the vulnerability indices (for surface or groundwater) is at its maximum (« $SRFWaterVul_g$ or $SOTWaterVul_g$ » equal to 3).

Between 2012 and 2018, the total potable water distributed in Corsica ($CleanWater_t$) averaged approximately 50 million m³ [43]. In this scenario, two levels of maximum potable water withdrawal thresholds for hydrogen production were considered: 0.1% and 0.05% of this total.

$$WATERCONS_{g,t,m} = \sum_i \frac{PT_{i,g,t,m} \times ElWUC}{1000} ; \forall g,t,m \qquad \text{*Equation 21}$$

$$WCV_{g,t,m} = WATERCONS_{g,t,m} \times dM_m \times WaterVulsn_{g,m} \qquad \text{*Equation 22}$$

$$WaterVul^{min}_{g,m} \times CleanWater_t \times minCW \times 1000000 \leq WCV_{g,t,m} \leq WaterVul^{max}_{g,m} \times CleanWater_t \times maxCW \times 1000000 ; \forall g,t,m ; \qquad \text{*Equation 23}$$

$$CleanWater_t = 50; WaterVul^{min}_{g,m} = 1; WaterVul^{max}_{g,m} = \{4;5\}; \forall g,t,m \qquad \text{*Equation 24}$$

### 4.2.3 Facility location and model parameters

In this study Qgis [14] was used to identify potential sites for hydrogen production and distribution facilities, as well as to estimate transport distances, following the approach in [7]. Among the 133 gas stations distributed across the territory, fifty-four were selected as candidate sites.

A consistent set of techno-economic, environmental, and risk parameters was then applied to production, storage, transport, and distribution components, as illustrated in Figure 1. The calculation methods and parameter values are aligned with those used in our previous work [7], and the complete parameter table is provided in the appendix.

## 5 Results and discussions

The developed model processes 1,070,275 simple equations before obtaining an optimal solution that considers all introduced constraints. These equations consist of both continuous variables (553,580) and discrete variables (147,600).

Table 4 presents a summary of optimization criterion values, including total system cost, overall GHG emissions, global risk index, and discounted production cost. These values are also shown for the optimal configuration resulting from the compromise among these three optimization criteria (multi-objective optimization) after applying the m-TOPSIS method.

| Key indicator / Objective functions | Total system cost (k€/day) | GHG emission (tCO$_2$e/day) ou (kg CO$_2$e/kg H$_2$) | Risk index | LCOH (€/kg) |
|---|---|---|---|---|
| **Minimizing system cost** | 61,4 | 22 (1,75) | 51,6 | **6,54** |
| **Minimizing GHG emission** | 84,8 | **16,6** (**1,32**) | 263,9 | 10,45 |
| **Minimizing risk index** | 76,9 | 21,3 (1,69) | **49,1** | 8,22 |
| **Multi-Objective optimization (MO) M-TOPSIS Criteria weigh : Equivalent** | **62,2** | 21,98 (1,75) | 53 | 6,55 |
| **Multi-Objective optimization (MO) M-TOPSIS Criteria weigh : Priority to GHG criteria** | 108 | 18 (1,43) | 54 | 11,4 |

*Table 4: Key indicator values according to the objective function used*

### 5.1 Minimizing system cost criteria

Minimizing system costs results in a hydrogen production chain starting with 7 units in 2026, expanding to 43 by 2050. Large-scale (5 MW) electrolyzers are alkaline and located in high-demand areas, with 7 alkaline units operational by 2050. These facilities are presented in figure 5. Storage capacity remains constant, with 9 medium-sized units distributed across 9 zones. The distribution network includes both mini and large refueling stations, with only zone 7 requiring a large station in 2026. Hydrogen exports increase from 1.2 t/day in 2026 to 2.2 t/day in 2050, with zone 1 supplying zone 5 and zone 6 supplying zone 4. Winter exports start at 0.5 t/day, rising to 0.9 t/day in summer.

GHG emissions rise from 10.1 tCO$_2$e/day in 2030 to 48 tCO$_2$e/day by 2050, averaging 22 tCO$_2$e/day. The risk index increases from 46.9 in 2030 to 60.3 in 2050, reflecting system expansion.

Operational costs for production, storage, and distribution rise from €4,934/day in 2026 to €25,632/day in 2050, while transport costs average €408/day in 2026 and €627/day in 2050. Electrolyzers operate for an average of 14.8 hours/day, with most zones running them for less than 10 hours/day.

Initial investments include €18.2 million for production, €11.3 million for distribution, and €2.7 million for transport. Compression units cost €213,000 each, with two units installed. The global LCOH over 25 years is €6.54/kg, increasing to €16.71/kg with liquid hydrogen transport due to additional costs for liquefaction, storage, and cryogenic transport.

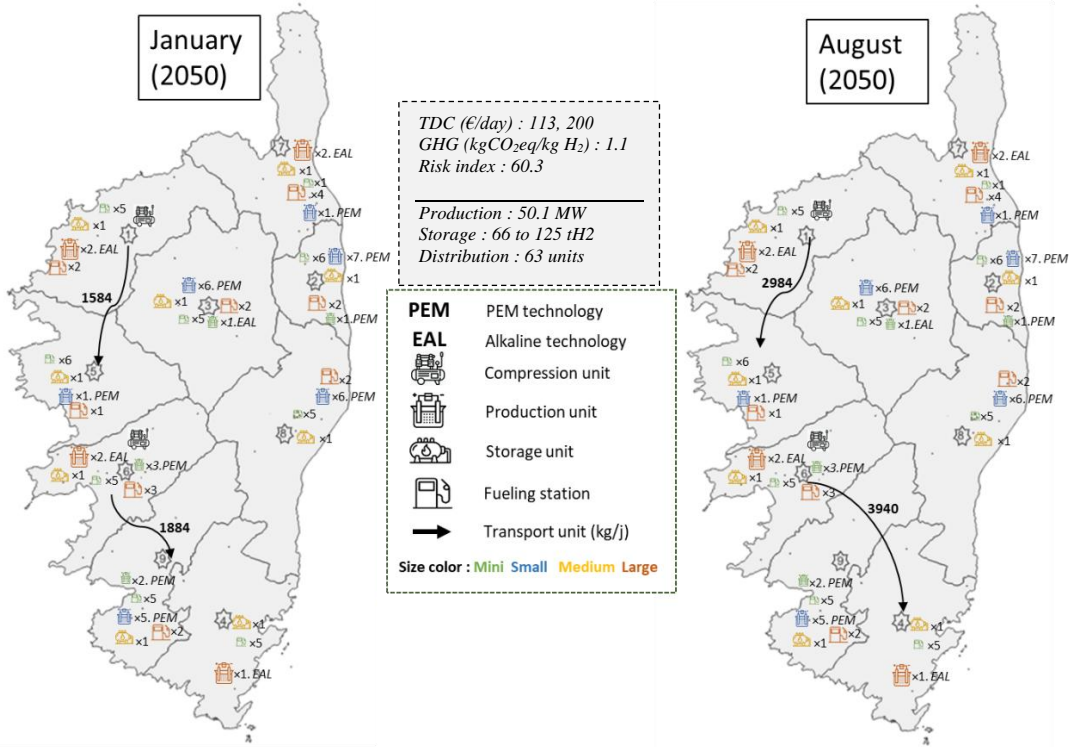

*Figure 5: Optimal HSC configuration in 2050 for minimal system cost*

### 5.1.1 Impact of Transportation Technology

According to some experts, the maintenance cost of FCEV is on average 20% to 30% lower than that of a combustion engine vehicle [34]. In this study, a 25% reduction in maintenance costs has been applied compared to a combustion engine vehicle.

For the use of conventional thermal transport trucks, the total investment cost for all distribution units is €1 million (and €1.72 million for liquid transport). Transitioning to retrofitted trucks reduces this cost to approximately €1.02 million. As for maintenance costs, they decrease from an average of €371/day to €361/day, respectively for thermal and electric (hydrogen retrofitted) transport. The LCOH resulting from this technology shift is around €6.44/kg (compared to €6.54/kg in thermal transport).

Additionally, there is an overall reduction in GHG emissions due to this change in transportation means. In total, approximately 1 ton of $CO_2$ equivalent emissions per day have been reduced, contributing to an average of 20.9 tons of $CO_2$ equivalent emissions per day across the entire distribution network.

### 5.1.2 Impact on water extraction restriction

By limiting the accessibility index for drinking water to 5 and setting the overall water extraction threshold to 0.1%, we observe an increase in water extraction in Zone 6 throughout the year, especially during the months of February, July, and August. However, unlike Zones 1, 4, and 7, Zone 6 is not under pressure in terms of drinking water accessibility (see figure 6). Regarding the areas under pressure, Zone 7 shows no changes in the quantity and distribution of water extraction. However, the amount of water extracted in Zone 1 decreases by nearly half, while in Zone 4 it increases slightly during January and July.

The increase in water extraction in Zone 6 has helped reduce water extraction in the pressured Zone 1 and maintained the extraction quantity in pressured Zone 7. These two zones represent key hydrogen production areas (along with another zone), unlike Zone 4, which begins production only from 2030 onward.

By reducing the extraction threshold from 0.1% to 0.05%, we observe a reduction in the amount of water extracted for Zone 6 with almost no change in the other zones. The extraction limit to meet hydrogen demand is reached.

Supply chain reorganization also occurs at the transport subsystem level. When water restrictions are considered, the amount of hydrogen transported (averaged over all periods) increases mainly during February and December, particularly from 2035 onwards. This indicates that when a certain demand level is reached, some importing zones shift a portion of their imports to months with fewer constraints for exporting zones. In return, these zones increase local production during these months (February and December) to meet local demand.

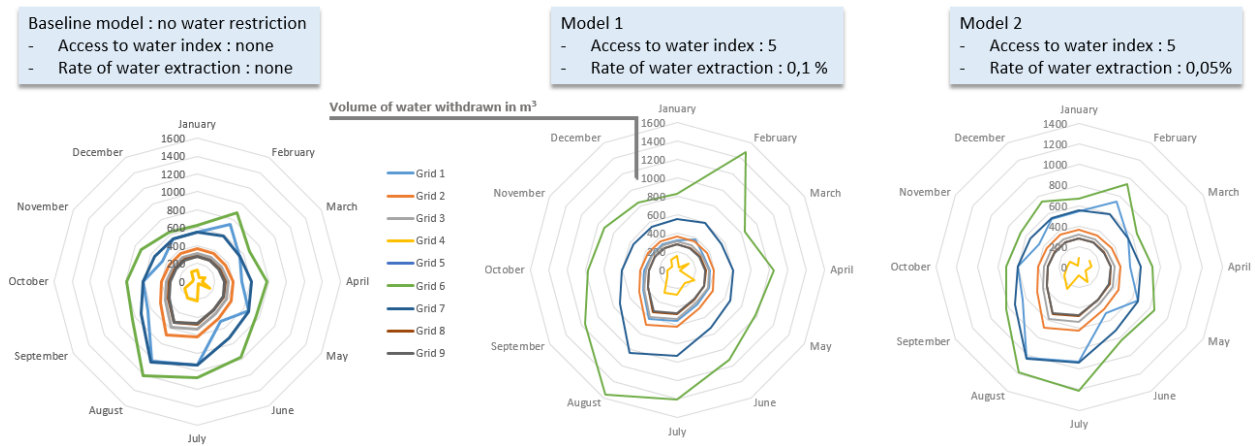

*Figure 6: Geographical and periodic distribution of water extraction for hydrogen production*

### 5.2 Minimizing GHG emission criteria

The majority of production units used in this configuration are "mini" sized at 300 kW. The configuration begins with 18 alkaline electrolyzers, of which 14 are mini-sized. Larger production units come into play starting from 2040, initially installed in Zone 1, followed by Zone 4. These facilities are shown in Figure 7.

This configuration is therefore "distributed," favoring a spread of production units across each zone.

In this setup, storage units are significantly oversized. A total of 213 storage units are installed from the first period onwards, all very small (50 kg).

The hydrogen supply chain (HSC) configuration includes 29 distribution units in 2026. All are very small, with the largest clusters in Zones 6 and 7, having 4 and 5 units respectively. Starting in 2040, three new small distribution stations will be required in Zone 7, and the following year, medium-sized units will be needed in Zones 6 and 7. Large-sized units will be necessary to meet demand by 2045.

There are also some hydrogen exports to Zones 4 and 5. Transport units serving these zones are present until 2040. After that, Zone 4 is no longer an importer, which reduces GHG emissions by favoring local production in Zone 4 after 2040.

Total GHG emissions are 7.9 tCO2e/day in 2030, rising to 35.8 tCO2e/day in 2050 (compared to 48 tCO2e/day at the same period when cost criteria are optimized). On average, global GHG emissions over the entire duration are about 16 tCO2e/day, distributed as follows: approximately 50% from the production subsystem, 40% from the storage subsystem, and 10% from the transport subsystem.

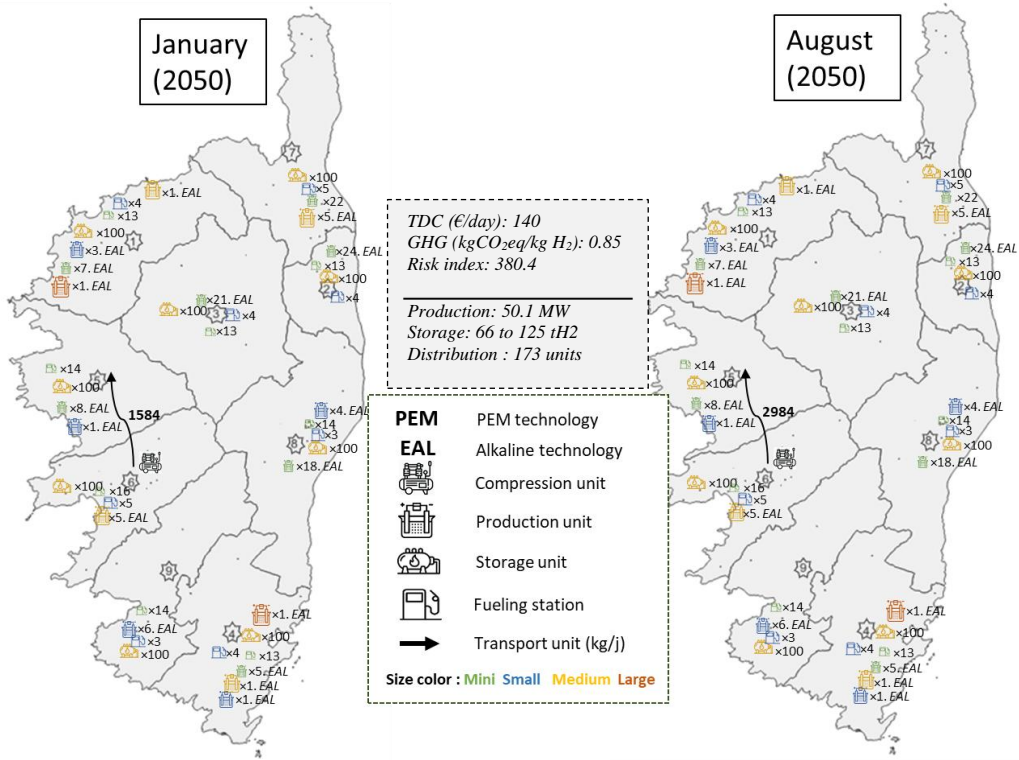

*Figure 7: Optimal HSC configuration in 2050 for minimal GHG emission*

The total risk index oevolves between 154 and 375 (respectively in 2030 and 2050). This overall risk index rate is approximately 2 to 3 times higher than that of the configuration obtained from the cost minimization criteria. On average, the production subsystems represent only 5% of the overall risk index, while transport and storage subsystems represent 10% and 85%, respectively. he use of small storage units contributes to the increase in the risk index.

Investment and maintenance costs are multiplied by nearly 2.5 compared to the cost minimization scenario: an initial investment of €58 million is required to implement this configuration, with approximately €110 million over the 25-year study period.

The maintenance and operation costs of the production, storage, and distribution subsystems are estimated at €5,135 per day in 2026, rising to €32,257 per day in 2050. Additional costs incurred by transport systems add up to €400 per day. The total operating costs of the entire system configuration (including one-time investments) are estimated at €84,800 per day on average over the entire study period.

## 5.3 Minimizing risk index criterion

In this configuration, there are 8 production units in 2026 (at the beginning of the study period). These 8 production units are diverse in terms of their nature and size. There are 5 small-sized units, including 2 PEM technology units and 3 of alkaline technology, and 3 medium-sized units, including 1 PEM technology unit and 2 alkaline technology units. All 9 zones are equipped with production units in 2026 except for Zones 4 and 5. These two zones are only equipped with production systems toward the end of the period (in 2049 and 2050). Additionally, no large-sized production units are observed throughout the system and during any periods. These facilities are presented in Figure 8.

The number of storage units is 9, with one medium-sized storage unit in each zone. The configuration of this sector is identical to that of the system obtained when optimizing the cost criterion. Therefore, this storage unit configuration is economically the most advantageous and reliable in terms of risk.

In 2026, at the distribution subsystem level, there are a total of 40 distribution units. All these units are small-sized ("mini") and remain so until the end of the study period.

However, Zone 5 hosts a significant number of distribution units despite having no production units in the short to medium term.

Total GHG emissions are 12.13 $tCO_2e$/day in 2030, rising to 41.7 $tCO_2e$/day in 2050. The former value was 35.8 $tCO_2e$/day and the latter 48 $tCO_2e$/day at the same periods for configurations optimized for GHG emissions and cost, respectively. On average, global GHG emissions are about 21 $tCO_2e$/day over the entire duration, representing a significant increase compared to the configuration minimizing GHG emissions, which emitted 16 $tCO_2e$/day. These GHG emissions are distributed as follows: approximately 70% associated with the production subsystem, 20% with the storage subsystem, and 10% with the transport subsystem.

In terms of investment, an initial investment of €34 million is required to implement this configuration. This corresponds to €24 million less compared to the configuration optimized for GHG emissions.

The maintenance and operation costs of the production, storage, and distribution subsystems are estimated at €5,756 per day in 2026, rising to €32,816 per day in 2050. The total operating costs of the entire system configuration (including one-time investments) are estimated at €7,690 per day on average over the entire study period.

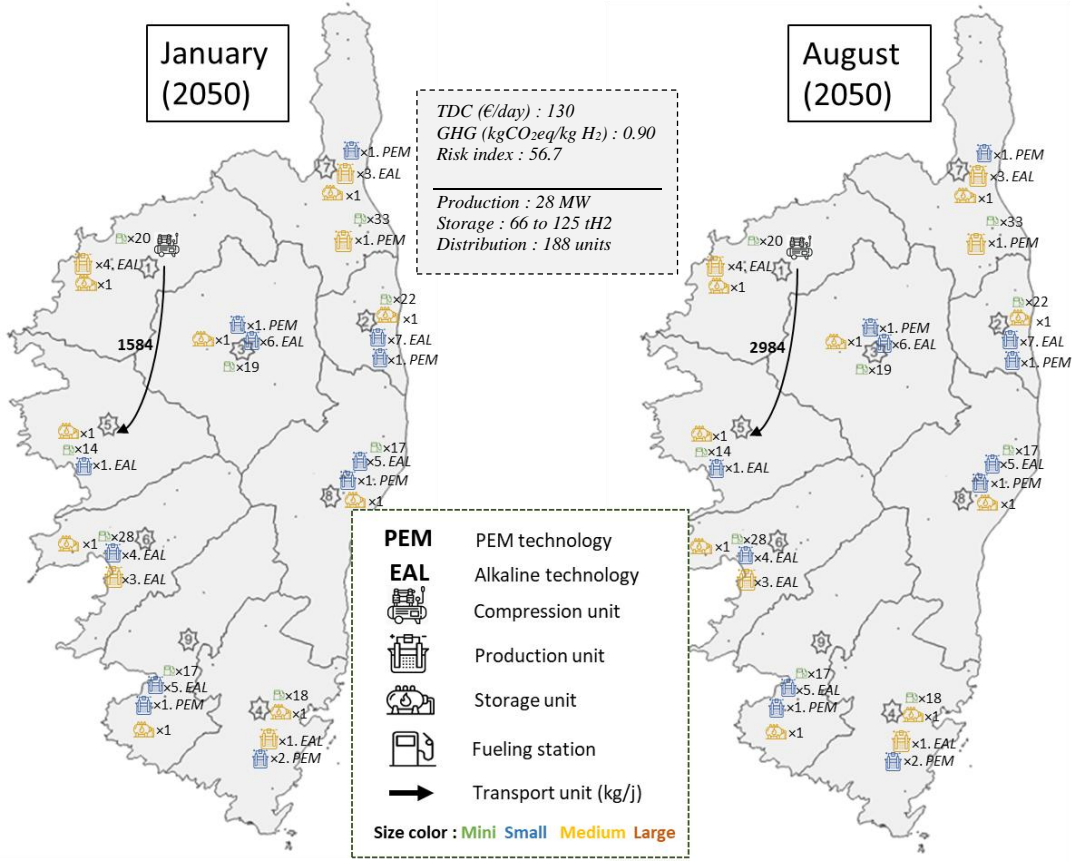

*Figure 8: Optimal HSC configuration in 2050 for minimal risk index*

## 5.4 Multi objective optimization

The simulation time of this model is longer compared to single-objective optimization models. Using the same computing hardware (Dell laptop, Core i5, 16 GB RAM), the model converges to an optimal result after more than 17 hours of simulation. Through the ε-constraint method, the model provides a set of solution points represented by the values of the three optimization criteria: total system cost (in k€/day), total GHG emissions (in t$CO_2$e/day), and the global risk index value. These solutions are presented in Figures 9a and 9b in the form of Pareto fronts, showing the relationship between cost and risk on one hand, and cost and GHG emissions on the other. The compromise solution from this Pareto front obtained after applying m-TOPSIS has a cumulative total system cost for 25 years of 1.55 M€/day (averaging to 62.2 k€/day for each year), a cumulative GHG emission of 549 t$CO_2$e/day (approximately 21.9 t$CO_2$e/day on average per year), and a cumulative global risk index over the study period of 1,325 (averaging a risk index value of 53).

The optimal result proposed using the m-TOPSIS method is depicted by the red triangles in Figure 9.

The optimal solution proposed by multi-objective optimization after applying m-TOPSIS with equal weights across criteria does not significantly reduce GHG emissions. The selected solution heavily leans towards minimizing cost-risk trade-off. Considering the importance of decarbonization, we decided to analyze optimal solutions from the Pareto front by adjusting the weight of GHG emissions using the m-TOPSIS method. Three other optimal solutions were identified based on the weight of the optimization criterion and the number of criteria studied. These solutions are presented in Figure 9.

When doubling the weight of the GHG emissions criterion (weights indicated in parentheses in Figure 9), we achieved a multi-objective solution with lower GHG emissions (18 tCO$_2$e/day). By excluding the criterion based on the risk index (Figure 9), we obtained optimal solutions with favorable GHG emissions (19 tCO$_2$e/day). At this level, two weights of the GHG emissions criterion were evaluated (weight=1 "black triangle" and weight=2 "purple triangle").

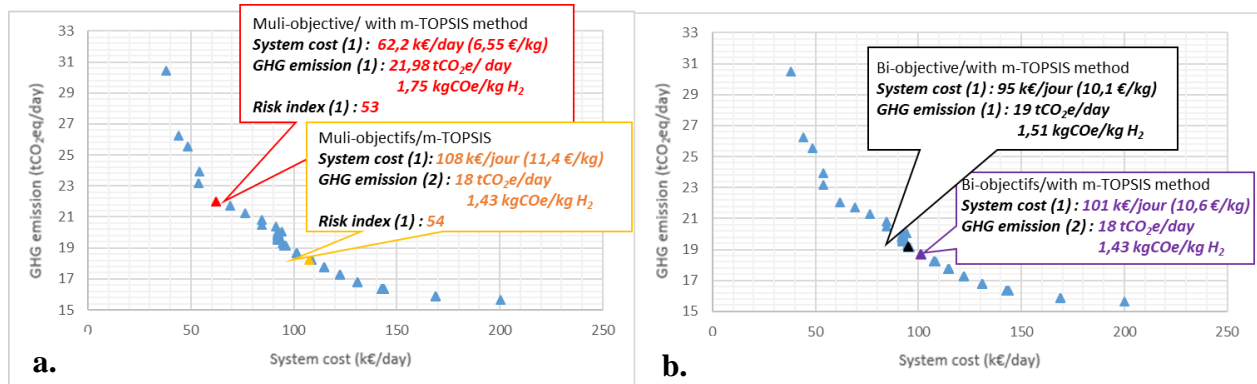

Figure 9: Multi objective and bi-objective optimization Pareto Front

The integration of water accessibility constraints has no significant impact on the optimal solutions of multi-objective optimization. A very slight decrease of approximately 0.7% is observed in the cost-optimal and risk-optimal solutions. There is no change in the optimal GHG emissions. The reorganization of the HSC solely results in a redistribution of production units.

# 6   Conclusion

The optimal configurations resulting from this study reveal a partially decentralized HSC, incorporating transport links to meet demand in areas where production is constrained by geographical limitations.

These efforts have identified a more balanced configuration of the HSC in terms of economics (LCOH of 6.55 €/kg), environmental impact (GHG emissions of 1.75 kgCO$_2$e/kg H2), and safety (global risk index of 54). Doubling the weight factor of the GHG criteria can improve the overall GHG emissions (down to 1.43 kg CO$_2$eq/kg H$_2$) but significantly increases the system cost (up to 10.6 €/kg). Therefore, the multi-objective optimization scenario with equal criterion weights set to 1 remains the most advantageous.

Moreover, the integration of water accessibility constraints, considering groundwater availability and seasonal precipitation variations, significantly influenced the spatial and temporal organization of hydrogen production. This led to a reallocation of production towards regions with stable water availability, such as Central Corsica, Pays Ajaccien, and Taravo/Valinco/Sartenais. However, the modeling could be further improved by considering the evolution of water resource conditions, which are impacted by both human interventions and environmental changes. Additionally, the assumption of constant water costs across territories could be revised to reflect local variations.

Furthermore, the use of retrofitted semi-trailer trucks as a means of transportation was evaluated. This alternative approach reduces GHG emissions by approximately 1 tCO$_2$e/day and decreases the overall hydrogen production cost by about 0.10 €/kg, primarily due to reduced transport unit maintenance costs. Additionally, it facilitates the deployment of heavy hydrogen vehicle fleets and distribution stations, thereby accelerating regional-scale hydrogen industry development.

Despite progress in HSC modeling, significant gaps remain in the explicit integration of potable water constraints. This study contributes to filling this gap by proposing a multi-objective, inter-seasonal MILP-based framework that incorporates water-energy nexus considerations, spatial-temporal discretization, and real-world constraints relevant to insular systems. On a policy level, it underlines the importance for regional authorities to integrate water resource governance into hydrogen strategies, particularly in alignment with the European Green Deal and EU Hydrogen Strategy. From a managerial perspective, it offers a practical decision-support tool for infrastructure planning and funding allocation, guiding investments toward high-impact, resource-efficient systems. On a theoretical level, the framework enriches the literature on sustainable energy transitions in resource-limited contexts. Future research should further extend this approach by developing decentralized hydrogen-water nexus models tailored to island environments and integrating seasonal dynamics in both energy and water supply. Such advances could ensure long-term resilience and sustainability of hydrogen deployment strategies in regions like Corsica.

# 7 Acknowledgements

This work has received research funding from the French government managed by the National Research Agency under the France 2030 program, with reference number ANR-22-EXES-0016.

# 8 Declaration of competing interest

The authors declare that they have no known competing financial interests or personal relationships that could have appeared to influence the work reported in this paper.

reseaux-et-stockage/1173-trajectoires-d-evolution-du-mix-electrique-a-horizon-2020-2060-9791029711732.html

# 10 Appendix

## A : Techno-economic model parameters

| | Capacity | CAPEX | OPEX | Efficiency | Number of sizes categories | Number of technologies type |
|---|---|---|---|---|---|---|
| **Production (Electrolyzer)** | 300-5 000 (kW) | 1038-3500 (€/kW) | 0.1-0.250 (€/kg) | 37.8-52 (kWh/kg) | 4 | 3 |
| **Conversion (Compressor)** | 126 (kg/h) | 1690 (€/kg) | 0.007 (€/kg) | 2.66 (kWh/kg) | 1 | 1 |
| **Conversion (Liquefier)** | 500 (kg/d) | 7,460 (€/kg) | - | 6.78 (kWh/kg) | 1 | 1 |
| **Storage (pressurized tank)** | 50-30 000 (kg) | 25-500 (€/kg) | 0.006-0.02 (€/kg) | - | 4 | 2 |
| **Distribution (Refueling station)** | 20-1,300 (kg/d) | 410-1,480 (k€) | 0.15-0.39 (€/kg) | 52.4-56.4 (kWh/kg) | 4 | 1 |
| **Transport (Truck)** | 670-4,300 (kg) | 746-200 (€/kg) | Depending on the distance travelled | - | 2 | 3 |
| **Electricity price from** | *ADEME Hypothesis 1 (A1) Reference trajectory* [44] | *ADEME Hypothesis 2 (A2) Nuclear extension* [44] | *PPE Hypothesis* | *EDF Green Tariff Hypothesis* | | |
| **Photovoltaic** | 0.014/kWh to €0.019/kWh between 2025 and 2050 | €0.031/kWh (2025), €0.034/kWh (2050) | €0.030/kWh | | | |
| **Wind** | €0.020/kWh from 2030, with a range of €0.029/kWh to €0.044/kWh between 2025 and 2050 | €0.031/kWh and €0.040/kWh between 2025 and 2050, with a low of €0.026/kWh in 2035 | €0.034/kWh | | | |
| **Hydropower** | Fixed at €0.010/kWh until 2050 | Fixed at €0.024/kWh until 2050 | | | | |
| **Grid electricity** | Between €0.047/kWh and €0.067/kWh for 2025 and 2050 | B€0.047/kWh and €0.068/kWh for 2025 and 2050 | €0.042/kWh | €0.1003/kWh (precisely €0.17/kWh on average in high season and €0.0306/kWh in low season | | |
| **Water cost** | | | 2.18 (€/m3) | | | |

## B : Retrofitted trucks parameters

|  | [a]**Tube trailer** | [b]**Tanker truck** |
|---|---|---|
| **Investment cost (k€)** | 510 | |
| **Maintenance cost** | - Driver wage: 20.47 €/hours<br>- Fuel consumption: 7.58 km/kg<br>  (13.2 kg/100 km)<br>- Fuel cost: 12 €/kg<br>- General expenses: 7.32 €/jour<br>- Load/unload time: 1.5 h[a]<br>  3 h[b]<br>- Maintenance cost: 0.03 €/km<br>- Average speed: 50 km/h<br>- Transport availability: 18h/day<br>- Transport weight: 40 tones<br>- GHG emission : 0 g$CO_2$/ton.km | |
| **Capacity (kg)** | 670[a] | 4300[b] |

## C : Tourist concentration by area

|  |  | Rate of tourist accommodation | Tourist attendance rate | Final rate of tourist visits |
|---|---|---|---|---|
| 1 | PAYS DE BALAGNE | 18% | 17% | 17,48% |
| 2 | CASTAGNICCIA / MARE E MONTI | 10% | 9% | 9,35% |
| 3 | CENTRE CORSE | 2% | 3% | 2,51% |
| 4 | EXTREME SUD / ALTA ROCCA | 24% | 26% | 24,95% |
| 5 | OUEST CORSE | 9% | 10% | 9,27% |
| 6 | PAYS AJACCIEN | 9% | 7% | 7,67% |
| 7 | PAYS BASTIAIS | 10% | 9% | 9,20% |
| 8 | PLAINE ORIENTALE | 11% | 11% | 10,87% |
| 9 | TARAVO / VALINCO / SARTENAIS | 8% | 9% | 8,69% |